\documentclass[twocolumn,epjc3]{svjour3}  

\smartqed  
\RequirePackage{graphicx}
\RequirePackage[numbers,sort&compress]{natbib}
\RequirePackage[colorlinks,citecolor=blue,urlcolor=blue,linkcolor=blue]{hyperref}
\journalname{Eur. Phys. J. C}

\usepackage{times}
\usepackage{float}
\usepackage{subfigure}
\usepackage{array}
\usepackage{amsmath}
\usepackage{amssymb}
\usepackage{amsfonts}
\usepackage{bm}
\usepackage{multirow}
\usepackage{booktabs}
\usepackage{dcolumn}
\usepackage{latexsym}
\usepackage{verbatim}
\usepackage{physics}
\usepackage[english]{babel}
\usepackage{widetext}
\usepackage[utf8]{inputenc}
\usepackage{todonotes}
\usepackage{epstopdf} 
\usepackage{microtype}

\usepackage{soul}

\graphicspath{{Figures/}}

\begin{document}

\title{Greybody factor for an electrically charged regular-de Sitter black holes in $d$-dimensions}

\author{Md Sabir Ali\thanksref{addr2,addr5,addr6,e1}, 
A. Naveena Kumara\thanksref{addr1,e2}, 
Kartheek Hegde\thanksref{addr1,e3}, 
C. L. Ahmed Rizwan\thanksref{addr1,addr3,e4}, 
Shreyas Punacha\thanksref{addr4,e5}, 
K. M. Ajith\thanksref{addr1,e6}}

\thankstext{e1}{e-mail: alimd.sabir3@gmail.com, sabir.pdf@iiserkol.ac.in}
\thankstext{e2}{e-mail: naviphysics@gmail.com}
\thankstext{e3}{e-mail: hegde.kartheek@gmail.com}
\thankstext{e4}{e-mail: ahmedrizwancl@gmail.com}
\thankstext{e5}{e-mail: shreyasp444@gmail.com, shreyas4@uw.edu}
\thankstext{e6}{e-mail: ajith@nitk.ac.in}

\institute{Department of Physics, National Institute of Technology Karnataka, Surathkal 575 025, India \label{addr1} 
\and Department of Physics, Indian Institute of Technology, Ropar, Rupnagar, Punjab 140 001, India \label{addr2} 
\and Department of Physical Sciences, Indian Institute of Science Education and Research Kolkata, Mohanpur, WB 741 246, India \label{addr5}
\and Department of Physics, Mahishadal Raj College, West Bengal, 721 628 India \label{addr6}
\and Department of Physics, Kannur University, Payyanur, Kerala, 670 327, India \label{addr3} 
\and Department of Oral Health Sciences, School of Dentistry, University of Washington, Seattle, WA 98195, USA \label{addr4}
}

\date{Received: date / Accepted: date}

\maketitle

\begin{abstract}
We investigate the propagation of scalar fields in the gravitational background of higher-dimensional, electrically charged, regular de Sitter black holes. Using an approximate analytical approach, we derive expressions for the greybody factor for both minimally and non-minimally coupled scalar fields. In the low-energy regime, we find that the greybody factor remains non-zero for minimal coupling but vanishes for non-minimal coupling, indicating a significant influence of curvature coupling on the emission profile. Examining the greybody factor alongside the effective potential, we explore how particle parameters (the angular momentum number and the non-minimal coupling constant) and spacetime parameters (the dimension, the cosmological constant, and the non-linear charge parameter) affect particle emission. While non-minimal coupling and higher angular momentum modes generally suppress the greybody factor, the non-linear charge parameter enhances it. We then compute the Hawking radiation spectra for these black holes and observe that, despite the non-linear charge enhancing the greybody factor, both non-minimal coupling and the non-linear charge ultimately reduce the total energy emission rate. These results provide insights into how modifications to classical black hole solutions in higher dimensions, through the inclusion of non-linear electrodynamics, impact their quantum emission properties.
\end{abstract}

\keywords{Greybody factor, Hawking radiation, Higher-dimensional black holes, Regular black holes, de Sitter spacetime, Non-linear electrodynamics}

\maketitle


\section{Introduction}

Black holes have long been a cornerstone in the exploration of gravitational physics, offering profound insights into the nature of spacetime, quantum mechanics, and thermodynamics. Classical solutions to Einstein's field equations, such as the Schwarzschild and Kerr black holes, predict the existence of singularities—points where spacetime curvature becomes infinite and the known laws of physics break down. These singularities not only challenge our theoretical frameworks but also highlight the limitations of classical general relativity. To address these profound issues, the concept of \emph{regular black holes} has emerged as a promising candidate. Unlike their classical counterparts, regular black holes are devoid of singularities, featuring a nonsingular core that ensures finite curvature invariants throughout spacetime.

The journey toward understanding regular black holes began with Bardeen's seminal work in 1968, where he proposed a static, spherically symmetric solution that intriguingly avoided singularities \cite{1968qtr..conf...87B}. Although Bardeen's solution was not an \emph{exact} solution to Einstein's equations, it laid the groundwork for future explorations into singularity-free black hole models. Nearly three decades later, Ayón-Beato and García made significant strides in this area by deriving exact regular black hole solutions within the framework of \emph{non-linear electrodynamics} (NLED) coupled to Einstein gravity. Their pioneering work, first published in \cite{Ayon-Beato:1998hmi}, introduced electrically charged regular black holes that possessed a proper Maxwell limit. They further expanded this framework by introducing diverse solutions through modifications of the NLED Lagrangian \cite{Ayon-Beato:1999qin, Ayon-Beato:1999kuh}.

Later, Ayón-Beato and García reinterpreted Bardeen's original solution as a non-linear magnetic monopole, thereby providing an \emph{exact} NLED-based solution for the Bardeen black hole \cite{Ayon-Beato:2000mjt}. Building on this foundation, subsequent studies extended the landscape of regular black holes by presenting the Bardeen model as a solution to Einstein's equations in the presence of an electric source \cite{Rodrigues_2018}. Additionally, higher-dimensional generalizations of the Bardeen black hole were introduced, further enriching the diversity of regular black hole models \cite{Ali:2018boy}. In recent years, a plethora of studies have emerged on regular black hole solutions and related topics \cite{K.:2020rzl,NaveenaKumara:2020lgq,A.:2019mqv,Ali:2019rjn,Ali:2019myr,Amir:2016cen,Abdujabbarov:2016hnw,Ghosh:2020ece,Jusufi:2020agr,Ghosh:2018bxg,Kumar:2018vsm,Ghosh:2020tgy,Amir:2020fpa,Ahmed:2020dzj,Ahmed:2020jic}. In this work, we present a class of higher-dimensional electrically charged regular solutions that include the Bardeen and Hayward cases.\footnote{Hayward proposed a framework that incorporates the dynamic formation, quiescence, and evaporation phases of nonsingular black holes, supported by physically reasonable energy conditions \cite{Hayward:2005gi}.}

The exploration of black hole solutions in higher-dimensi\-onal spacetimes has garnered substantial interest, not only due to their applications in string theory and braneworld scenarios but also because they offer intrinsic insights into the nature of gravity and spacetime. Studying General Relativity in more than four dimensions allows physicists to investigate how fundamental properties of black holes extend beyond the familiar four-dimensional case, revealing unique features, such as altered thermodynamic behaviour, modified stability criteria, and the possibility of non-spherical horizon topologies \cite{Emparan:2008eg}. Understanding these differences helps determine which characteristics are peculiar to four dimensions and which are general properties of gravitational theory. Additionally, higher-dimensional black holes play a crucial role in the AdS/CFT correspondence, relating gravitational phenomena in \(d\)-dimensional Anti-de Sitter space to quantum field theories in \(d-1\) dimensions \cite{Aharony:1999ti}. They also have implications for high-energy physics, where scenarios involving large extra dimensions and TeV-scale gravity suggest the conceivable production of microscopic black holes in future colliders \cite{Cavaglia:2002si, Kanti:2004nr}. As mathematical objects, these black hole spacetimes are among the most significant Lorentzian Ricci-flat manifolds in any dimension, and their study contributes to a deeper understanding of classical black holes and the extremal behaviour of spacetime.

Incorporating a positive cosmological constant (\(\Lambda\)) introduces a de Sitter (dS) background, which not only modifies the asymptotic structure of spacetime but also introduces a cosmological horizon alongside the event horizon of the black hole. This modification has profound implications for the propagation of fields and particles, as well as for the thermodynamic properties and stability of black holes \cite{Pappas:2017kam, Kanti:2017ubd,PhysRevD.55.7538, Kanti:2014dxa}.

A critical aspect of black hole physics is the phenomenon of Hawking radiation \footnote{The consideration of solutions in extra spacetime dimensions has provided a significant boost in this direction, as these dimensions may lead to the formation of mini black holes in high-energy particle colliders \cite{Giddings_2002, PhysRevLett.87.161602, Landsberg_2003} or cosmic ray interactions \cite{PhysRevLett.88.021303, PhysRevD.65.064023, PhysRevD.65.124027, Ringwald_2002}, enabling the potential detection of associated Hawking radiation \cite{Kanti:2014vsa,PhysRevD.80.084016}. Notably, Hawking radiation is also linked to quasinormal modes \cite{_vg_n_2018,catalan2015quasinormal,Devi:2020uac, Dey:2018cws, Rinc_n_2020} as well as black hole shadows \cite{Jusufi:2020mmy, Jusufi:2020agr, PhysRevD.101.084055}.}, wherein black holes emit particles due to quantum effects near the event horizon \cite{Hawking:1974sw}. This radiation is inherently thermal; however, the spectrum deviates from that of a perfect blackbody due to the presence of a greybody factor \cite{Unruh:1976fm, Page:1976df}. The greybody factor quantifies the probability that emitted particles can traverse the potential barriers surrounding the black hole and reach an observer at infinity. Understanding the greybody factor is essential for accurately characterizing the emission spectra and for probing the underlying spacetime geometry \cite{Harmark:2007jy} (see Ref \cite{Andersson:2000tf} for an interesting read). 

In this study, we investigate the propagation of scalar fields in the gravitational background of higher-dimensional electrically charged regular Bardeen-de Sitter black holes. Scalar fields, governed by the Klein-Gordon equation, serve as a simplified yet insightful model for exploring wave propagation in curved spacetimes \cite{Andersson:2000tf}. The complexity of the spacetime geometry necessitates the use of approximate analytical methods to solve the radial part of the scalar field equation. We employ a matching technique, solving the radial equation in distinct regions—near the event horizon and near the cosmological horizon—and subsequently matching these solutions in an intermediate region \cite{Crispino:2013pya, Kanti:2014dxa}. This approach facilitates the derivation of an analytical expression for the greybody factor in the low-energy regime.

Our analysis reveals that both spacetime properties, such as the cosmological constant $\Lambda$, the non-linear charge parameter $q$, and the spacetime dimension $d$, as well as particle properties, including the angular momentum number $l$ and the non-minimal coupling constant $\xi$, significantly influence the greybody factor. Specifically, higher angular momentum numbers and stronger non-minimal couplings tend to suppress the greybody factor, thereby reducing the probability of low-energy particle emission. Conversely, the non-linear charge parameter $q$ enhances the greybody factor by lowering the effective potential barrier, facilitating greater emission probabilities. The cosmological constant exhibits a nuanced dual role: it can either enhance or suppress the greybody factor depending on the value of the coupling constant $\xi$, highlighting a complex interplay between scalar field coupling and cosmological expansion.

Furthermore, we compute the power spectra of Hawking radiation for these black holes, which depend intricately on both the greybody factor and the black hole's temperature. The temperature itself is influenced by the non-linear charge and the cosmological constant, leading to a rich tapestry of dependencies in the emission rates. Our results indicate that non-minimal coupling generally suppresses the total energy emission rate, while the effects of the non-linear charge are multifaceted due to its simultaneous influence on the greybody factor and the black hole temperature. The cosmological constant continues to play a dual role in energy emission, consistent with its behaviour in influencing the greybody factor.

This investigation extends the understanding of regular black holes in higher-dimensional de Sitter spacetimes and provides valuable insights into how modifications to classical black hole solutions affect quantum processes like the Hawking radiation. The findings hold potential implications for theories involving extra dimensions and contribute to the ongoing efforts to reconcile general relativity with quantum mechanics. Moreover, the study of greybody factors in such rich geometries opens avenues for future research, including numerical analyses to explore the high-energy regime and extensions to other types of fields, such as spinor and vector fields.

The paper is organized as follows: In Section~2, we present the derivation of electrically charged regular black hole solutions in \(d\)-dimensional de Sitter spacetime, discussing their properties and physical interpretation. Section~3 focuses on the formulation of the Klein-Gordon equation in this background and the challenges associated with solving it. In Section~4, we derive an approximate analytical expression for the greybody factor by matching solutions near the event and cosmological horizons. Section~5 analyses the effects of various spacetime and particle parameters on the greybody factor and the effective potential, providing insights into the emission of scalar particles. In Section~6, we compute the power spectra of Hawking radiation, exploring how the parameters influence the energy emission rate. Finally, in Section~7, we summarize our findings and discuss potential avenues for future research, including numerical methods to address the high-energy regime and extensions to other types of fields.

\section{Regular-de Sitter Black Holes in \(d\) Dimensions}\label{section2}

In this section, we briefly present the exact solution of a regular black hole in a \(d\)-dimensional de Sitter background \cite{Ali:2018boy}. We start with the action describing general relativity coupled to non-linear electrodynamics in \(d\) dimensions,
\begin{equation}
\label{action-d}
S = \frac{1}{16\pi} \int d^{d}x \sqrt{-g} \left[ R + 2\Lambda - \mathcal{L}(\mathcal{F}) \right],
\end{equation}
where \(d\) is the spacetime dimension, \(R\) is the Ricci scalar, \(\Lambda\) is the cosmological constant, and the Lagrangian density \(\mathcal{L}(\mathcal{F})\) is a function of \(\mathcal{F} = F_{\mu\nu} F^{\mu\nu}\). Here, \(F_{\mu\nu} = \nabla_{\mu} A_{\nu} - \nabla_{\nu} A_{\mu}\) is the electromagnetic field strength tensor. The variation of the action \eqref{action-d}, \(\delta S = 0\), yields the equations for the gravitational field and the non-linear Maxwell Field,
\begin{align}
\label{einstein}
R_{\mu\nu} - \frac{1}{2} g_{\mu\nu} R &= T_{\mu\nu},\\
\nabla_{\mu} \left( \mathcal{L}_{\mathcal{F}} F^{\mu\nu} \right) &= 0, \quad \nabla_{\mu} \left( \ast F^{\mu\nu} \right) = 0,
\end{align}
where \(\mathcal{L}_{\mathcal{F}} = \partial \mathcal{L} / \partial \mathcal{F}\), and the energy-momentum tensor is given by
\begin{equation}
\label{emt-1}
T_{\mu\nu} = 2 \left[ \mathcal{L}_{\mathcal{F}} F_{\mu\alpha} F_{\nu}^{\ \alpha} - \frac{1}{4} g_{\mu\nu} \mathcal{L}(\mathcal{F}) \right].
\end{equation}

Using the Einstein and Maxwell equations, it was shown that a four-dimensional asymptotically flat Bardeen black hole can be associated with an electric source \cite{Rodrigues_2018}. Here, we extend the argument to the more general case in arbitrary \(d\) dimensions with a positive cosmological constant \(\Lambda\) for a family of regular black hole solutions described by the action \eqref{action-d}. Considering a static and spherically symmetric black hole, the only non-zero component of the electromagnetic field tensor is \(F^{10}\). From Maxwell's equations, we have
\begin{equation}
F^{10} = \frac{q^{d-3}}{r^{d-2}} \mathcal{L}_{\mathcal{F}}^{-1}.
\end{equation}

The metric for a class of spherically symmetric, electrically charged regular de Sitter black holes is given by
\begin{equation}
\label{eq:4metric}
ds^{2} = -h(r)\, dt^{2} + \frac{dr^{2}}{h(r)} + r^{2} d\Omega_{d-2}^{2},
\end{equation}
where the metric function is
\begin{equation} \label{heqn}
h(r) = 1 - \frac{m(r)}{r^{\alpha - 2}} - \frac{2\Lambda r^{2}}{(d - 1)(d - 2)},
\end{equation}
and the mass function is
\begin{equation} \label{meqn}
m(r) = \frac{\mu\, r^{\alpha}}{\left( r^{\beta} + q^{\beta} \right)^{\alpha / \beta}}.
\end{equation}
The metric function reduces to the Bardeen solution when \(\alpha = d - 1\) and \(\beta = d - 2\), while for \(\alpha = \beta = d - 1\), it reduces to the Hayward solution. The parameter \(\mu\) is related to the mass of the black hole \(M\) by
\begin{equation}
\mu = \frac{16\pi G M}{(d - 2) \Omega_{d - 2}},
\end{equation}
where \(\Omega_{d - 2}\) is the volume of the unit \((d - 2)\)-sphere. Here, \(q\) is the non-linear parameter arising when gravity is coupled to non-linear electrodynamics (in our case, \(q\) is interpreted as the electric charge). In the limit \(q \to 0\), the metric reduces to that of the Schwarzschild–de Sitter black hole. For the electric case, from Eqs.~\eqref{einstein} and \eqref{emt-1}, we can write the non-zero components of the Einstein equations as,
\begin{align}
\label{lagrangian-1}
\frac{\alpha\, m'(r)}{r^{2}} &= \mathcal{L} + \frac{q^{2(d - 3)}}{r^{2(d - 2)}}\, \mathcal{L}_{\mathcal{F}}^{-1}, \\
\label{lagrangian-2}
\frac{m''(r)}{r} &= \mathcal{L}.
\end{align}
In terms of the horizon radius \(r_{h}\), \(\mu\) can be expressed as
\begin{equation} \label{mueqn}
\mu = \left( 1 + \frac{q^{\beta}}{r_{h}^{\beta}} \right)^{\alpha / \beta} r_{h}^{\alpha - 2} \left( 1 - \frac{2\Lambda r_{h}^{2}}{(d - 1)(d - 2)} \right).
\end{equation}

The constant \(q\) has a significant effect on the stability of regular black holes. Through perturbation analysis, it was found that regular de Sitter black holes are unstable in certain parameter regions \cite{Fernando:2016ksb, Dey:2018cws, Wahlang:2017zvk}. In our discussions, we restrict the parameters to the stable region such that spacetime always has two horizons: the black hole horizon \(r_{h}\) and the cosmological horizon \(r_{c}\).

The Lagrangian \(\mathcal{L}(r)\) and its derivative \(\mathcal{L}_{\mathcal{F}}(r)\) are obtained by solving Eqs.~\eqref{lagrangian-1} and \eqref{lagrangian-2}, yielding
\begin{align}
\mathcal{L}(r) &= \frac{\mu\, r^{\alpha - 3} q^{\beta} \left[ (\alpha - 1)\alpha q^{\beta} - (\beta + 1)\alpha r^{\beta} \right]}{\left( r^{\beta} + q^{\beta} \right)^{(\alpha + 2\beta) / \beta}}, \\
\mathcal{L}_{\mathcal{F}}(r) &= \frac{\left( r^{\beta} + q^{\beta} \right)^{(\alpha + 2\beta) / \beta}}{\mu\, \alpha (2\beta + 1)\, r^{\alpha + \beta + 1 - d}\, q^{\beta - 2(d - 3)}}.
\end{align}
For \(\alpha = d - 1\) and \(\beta = d - 2\), we obtain \(\mathcal{L}(r)\) and its derivative \(\mathcal{L}_{\mathcal{F}}(r)\) for the Bardeen case,
\begin{align}
&\mathcal{L}(r) = \frac{\mu\, r^{d - 4} q^{d - 2} \left[ (d - 2)(d - 1) q^{d - 2} - (d - 1)^{2} r^{d - 2} \right]}{\left( r^{d - 2} + q^{d - 2} \right)^{(3d - 5) / (d - 2)}}, \\
&\mathcal{L}_{\mathcal{F}}(r) = \frac{q^{d - 4} \left( r^{d - 2} + q^{d - 2} \right)^{(3d - 5) / (d - 2)}}{\mu (d - 1)(2d - 3) r^{2(d - 1)}}.
\end{align}
When \(d = 4\), the above expressions for the Lagrangian and its derivative reduce to those given in Ref.~\cite{Rodrigues_2018}. For \(\alpha = \beta = d - 1\), we obtain \(\mathcal{L}(r)\) and its derivative \(\mathcal{L}_{\mathcal{F}}(r)\) for the electrically charged Hayward black hole as
\begin{align}
&\mathcal{L}(r)= \frac{\mu\, r^{d - 4} q^{d - 1} \left[ (d - 2)(d - 1) q^{d - 1} - d(d - 1) r^{d - 1} \right]}{\left( r^{d - 1} + q^{d - 1} \right)^{3}}, \\
&\mathcal{L}_{\mathcal{F}}(r) = \frac{q^{d - 5} \left( r^{d - 1} + q^{d - 1} \right)^{3}}{\mu (d - 1)(2d - 1) r^{2d - 1}}.
\end{align}
Although we have a family of electrically charged regular black hole solutions for different combinations of \(\alpha\) and \(\beta\), in the subsequent analysis, we focus only on the electrically charged regular Bardeen–de Sitter black holes.

As a consistency check and for the physical relevance of the solutions and the source, we examine the energy conditions. We identify the terms \(T^{t}_{\ t} = -\rho\), \(T^{r}_{\ r} = p_{r}\), and \(T^{\theta_{i}}_{\ \theta_{i}} = p_{t}\) (for \(i = 1, 2, \ldots, d - 2\)), where \(\rho\), \(p_{r}\), and \(p_{t}\) are the energy density, radial pressure, and tangential pressure, respectively. The energy conditions are given by~\footnote{SEC: strong energy condition, WEC: weak energy condition, DEC: dominant energy condition, NEC: null energy condition}
\begin{align}
\label{energy-cond1}
\text{SEC}(r) & = \quad \rho + \sum_{i = 1}^{d - 1} p_{i} = (d - 2) p_{t} \geq 0, \nonumber \\
\text{WEC}_{1,2}(r) & = \quad \text{NEC}_{1,2}(r) = \rho + p_{r,t} \geq 0, \nonumber \\
\text{WEC}_{3}(r) & = \quad \text{DEC}_{1}(r) = \rho \geq 0, \nonumber \\
\text{DEC}_{2,3}(r) & = \quad \rho - p_{r,t} \geq 0.
\end{align}

\begin{figure*}[ht!]
\centering
\includegraphics[width=\textwidth]{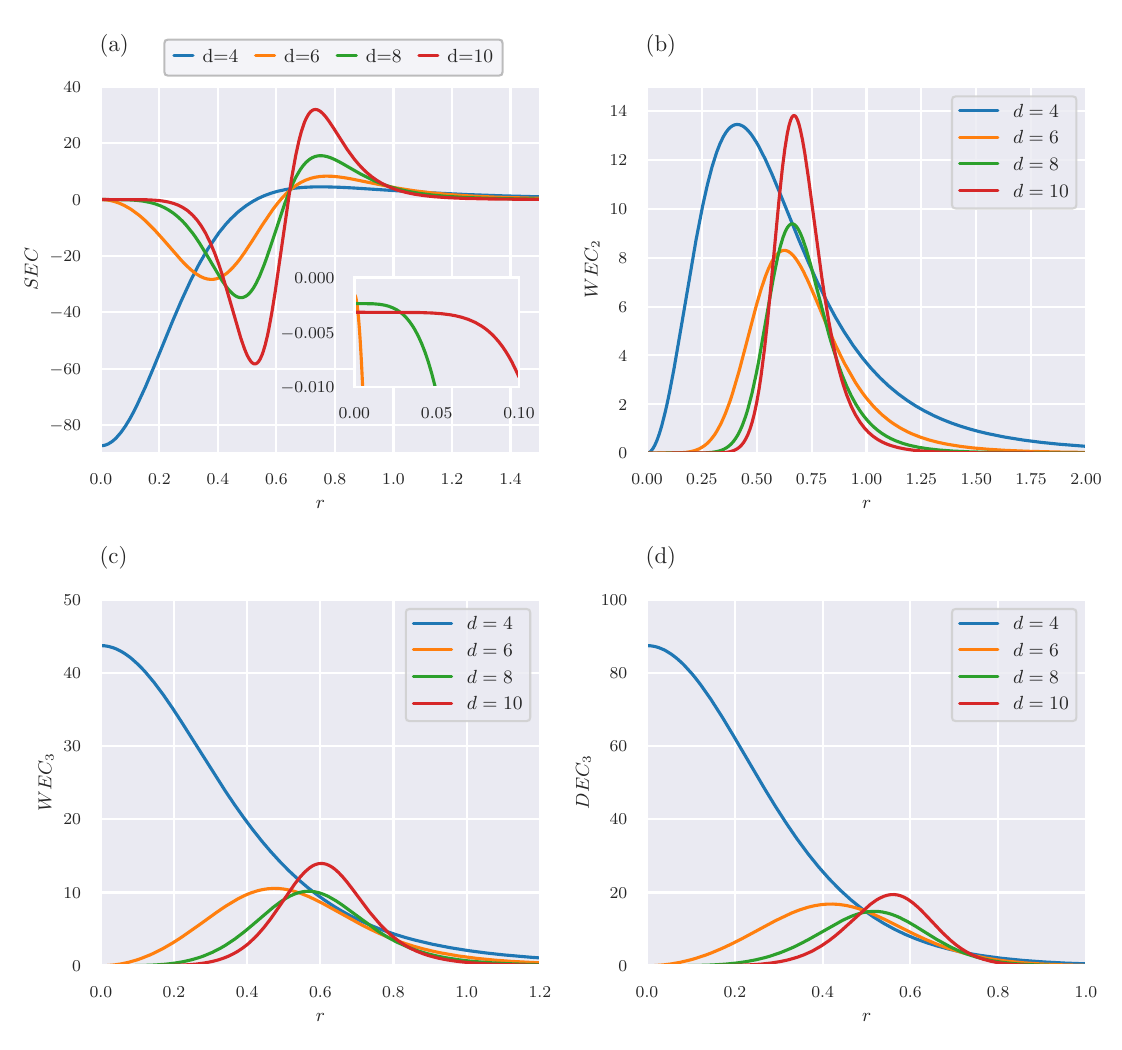}
\caption{Energy conditions $SEC$, $WEC_2$, $WEC_3$, and $DEC_3$. As expected for a regular black hole, the $SEC$ is violated for arbitrary $d$ dimensions. The behavior of $SEC$ near the center is shown in the inset of (a). In the plots, we set $\Lambda = 0.01$, $q = 0.65$, and $M = 1$. (Since $DEC_2 = 2 \times WEC_3$, only $WEC_3$ is shown.)}
\label{fig0}
\end{figure*}

We calculate the expressions for the energy density, radial pressure, and tangential pressure as
\begin{align}
\label{rho-pp}
\rho(r) = & \frac{\mu (d - 1)(d - 2) r^{d - 4} q^{d - 2}}{\left( r^{d - 2} + q^{d - 2} \right)^{(2d - 3)/(d - 2)}} + \frac{\Lambda}{8\pi}, \nonumber \\
p_{r}(r) = & -\frac{\mu (d - 1)(d - 2) r^{d - 4} q^{d - 2}}{\left( r^{d - 2} + q^{d - 2} \right)^{(2d - 3)/(d - 2)}} - \frac{\Lambda}{8\pi}, \nonumber \\
p_{t}(r) = & \frac{\mu r^{d - 4} q^{d - 2} \left[ - (d - 2)(d - 1) q^{d - 2} + (d - 1)^{2} r^{d - 2} \right]}{\left( r^{d - 2} + q^{d - 2} \right)^{(3d - 5)/(d - 2)}} \nonumber \\ & - \frac{\Lambda}{8\pi}.
\end{align}
Therefore, the expressions for the energy conditions are
\begin{widetext}
\begin{align}
\label{energy-cond}
\text{SEC}(r) &= (d - 2) \left( \frac{\mu r^{d - 4} q^{d - 2} \left[ - (d - 2)(d - 1) q^{d - 2} + (d - 1)^{2} r^{d - 2} \right] }{ \left( r^{d - 2} + q^{d - 2} \right)^{(3d - 5)/(d - 2)} } - \frac{\Lambda}{8\pi} \right), \nonumber \\
\text{WEC}_{1}(r) &= 0, \nonumber \\
\text{WEC}_{2}(r) &= \frac{\mu (d - 1)(2d - 3) r^{2(d - 3)} q^{d - 2} }{ \left( r^{d - 2} + q^{d - 2} \right)^{(3d - 5)/(d - 2)} }, \nonumber \\
\text{WEC}_{3}(r) &= \text{DEC}_{1}(r) = \frac{\mu (d - 1)(d - 2) r^{d - 4} q^{d - 2} }{ \left( r^{d - 2} + q^{d - 2} \right)^{(2d - 3)/(d - 2)} } + \frac{\Lambda}{8\pi}, \nonumber \\
\text{DEC}_{2}(r) &= \frac{2 \mu (d - 1)(d - 2) r^{d - 4} q^{d - 2} }{ \left( r^{d - 2} + q^{d - 2} \right)^{(2d - 3)/(d - 2)} } + \frac{\Lambda}{4\pi}, \nonumber \\
\text{DEC}_{3}(r) &= \frac{\mu (d - 1) r^{d - 4} q^{d - 2} \left[ 2(d-2)\, q^{d - 2} - r^{d - 2} \right] }{ \left( r^{d - 2} + q^{d - 2} \right)^{(3d - 5)/(d - 2)} } + \frac{\Lambda}{4\pi}.
\end{align}
\end{widetext}
The expressions \eqref{rho-pp} and \eqref{energy-cond} reduce to the corresponding expressions for the electrically charged Bardeen black holes in four dimensions when \(\Lambda = 0\) \cite{Rodrigues_2018}. From the above expressions, we observe that the strong energy condition is violated when \(r < \left( \frac{d - 2}{d - 1} \right)^{1 / (d - 2)} q\), which reduces to \(r < \left( \frac{2}{3} \right)^{1/2} q\) for \(d = 4\) in the limit \(\Lambda = 0\) \cite{Rodrigues_2018}. This is expected, since this region is inside the event horizon and the metric function becomes negative at this value of \(r\), a typical behaviour for any black hole spacetime with a regular centre. However, the condition \(p_{r} = p_{t}\) is not met by the \(d\)-dimensional electrically charged regular black holes, indicating an anisotropic perfect fluid with \(p_{r} = -\rho\). Therefore, $DEC_2=2 \times  WEC_3$. Following \cite{Rodrigues_2018} (see also \cite{Rodrigues:2025plw, Lobo:2020ffi, Junior:2020zdt}) we present the graphical representations of the energy conditions, as shown in Fig. \ref{fig0}.

\section{The Klein-Gordon Equation} \label{section3}

To explore the concepts involved in studying the scattering of waves by black holes, we consider the relatively simple case of scalar waves. Although no massless scalar fields have been observed in nature, scalar waves provide a useful model because the fundamental equations governing weak electromagnetic fields or gravitational waves in curved spacetime are essentially similar to the scalar field wave equation \cite{Andersson:2000tf}. Therefore, analysing scalar fields allows us to gain insight into the scattering processes. We assume that the higher-dimensional regular black hole emits scalar particles that carry small amounts of mass or energy compared to the black hole mass, so the gravitational background remains effectively unchanged \cite{Kanti:2014dxa}.

In this section, we study the emission of scalar particles that are minimally or non-minimally coupled to gravity. For the coupling of the scalar field \(\Phi\), we have an additional term in the action,
\begin{equation}
S_{\Phi} = -\frac{1}{2} \int d^{d}x \sqrt{-g} \left[ \xi \Phi^{2} R + \partial_{\mu} \Phi \partial^{\mu} \Phi \right],
\end{equation}
where \(\xi\) is the non-minimal coupling parameter determining the strength of the coupling; \(\xi = 0\) corresponds to minimal coupling. The propagation of the scalar particle is governed by the following equation of motion,
\begin{equation} \label{eqnofmotion}
\nabla_{\mu} \nabla^{\mu} \Phi = \xi R \Phi.
\end{equation}

It is customary to make an ansatz in a spherically symmetric background,
\begin{equation} \label{ansatz}
\Phi = e^{-i\omega t} \phi(r) Y_{(d-2)}^{l}(\Omega),
\end{equation}
where \(Y_{(d-2)}^{l}(\Omega)\) are the spherical harmonics on \(S^{d-2}\). Substituting the ansatz \eqref{ansatz} into the equation of motion \eqref{eqnofmotion}, we obtain a separable form, from which the radial equation reads
\begin{equation} \label{eq:RadialEq}
\frac{1}{r^{d-2}} \frac{d}{dr} \left( h r^{d-2} \frac{d\phi}{dr} \right) + \left[ \frac{\omega^{2}}{h} - \frac{l(l + d - 3)}{r^{2}} - \xi R \right] \phi = 0.
\end{equation}
Redefining the radial function as \(u(r) = r^{\frac{d-2}{2}} \phi(r)\), and introducing the tortoise coordinate \(r_{*}\) defined by \(dr_{*} = dr / h(r)\), we have
\begin{equation}\label{schrodingereqn}
\frac{d^{2}u}{dr_{*}^{2}} + \left( \omega^{2} - V(r) \right) u = 0,
\end{equation}
where the effective potential is
\begin{equation}
\begin{aligned}
V(r) =h(r) & \left[ \frac{l(l + d - 3)}{r^{2}} + \xi R + \frac{d - 2}{2r} h' \right. \\ & \left. \, + \frac{(d - 2)(d - 4)}{4r^{2}} h \right].
\end{aligned}
\end{equation}

The effective potential vanishes at the two horizons because the metric function \(h\) also vanishes there. The dependence of the effective potential on spacetime properties \((\Lambda, q, d)\) and particle properties \((\xi, l)\) can be studied from the above equation. In our analysis, we use equations \eqref{heqn} and \eqref{meqn}, setting \(r_{h} = 1\). The Ricci scalar is given by
\begin{widetext}
\begin{equation}
\begin{aligned}
R = \frac{\left( q^{\alpha} + r^{\beta} \right)^{ -\frac{\alpha}{\beta} - 2 }}{d^{2} - 3d + 2}  &\Bigg[  r^{2\beta} \left\{ \left( \alpha^{2} - 7\alpha + 12 \right) \left( d^{2} - 3d + 2 \right) \mu + 24 \Lambda \left( q^{\alpha} + r^{\beta} \right)^{\alpha / \beta} \right\} \\
&\quad +\; q^{\alpha} r^{\beta} \left\{ 48 \Lambda \left( q^{\alpha} + r^{\beta} \right)^{\alpha / \beta} - \left( d^{2} - 3d + 2 \right) \mu \left[ \alpha (\beta + 7) - 24 \right] \right\} \\
&\quad +\; 12 q^{2\alpha} \left\{ \left( d^{2} - 3d + 2 \right) \mu + 2 \Lambda \left( q^{\alpha} + r^{\beta} \right)^{\alpha / \beta} \right\} \Bigg],
\end{aligned}
\end{equation}    
\end{widetext}
where \(\mu\) is given by equation \eqref{mueqn}. We will illustrate these behaviours along with the profile of greybody factor in the following sections, as this will provide intuitive insight into particle propagation.

\section{Greybody Factor}\label{section4}

Hawking's seminal work \cite{Hawking:1974sw} demonstrated that black holes emit radiation with a thermal spectrum, using a semi-classical approximation. The expected number $\langle n(\omega) \rangle$ of particles of a given species emitted at frequency $\omega$ is given by
\begin{equation}\label{n(omega)}
\langle n(\omega) \rangle = \frac{\gamma(\omega)}{e^{\omega / T_H} \pm 1},
\end{equation}
where $T_H$ is the Hawking temperature, and the plus (minus) sign corresponds to fermions (bosons). The function $\gamma(\omega)$ is the \textit{greybody factor}, representing the probability that an outgoing wave with frequency $\omega$ will reach infinity. This factor is crucial because it introduces deviations from the perfect blackbody spectrum; if $\gamma(\omega)$ were constant (specifically, equal to one), the emission spectrum would be that of an ideal blackbody.

In terms of the Schrödinger-like problem \eqref{schrodingereqn}, the absorption probability $\gamma(\omega)$ can be expressed as $\gamma(\omega) = |T(\omega)|^2$, where $T(\omega)$ is the transmission coefficient for the potential barrier in the spacetime geometry under consideration. Thus, calculating the greybody factor amounts to determining the tunnelling probability through the effective potential barrier. Our goal is to compute $\gamma(\omega)$ for the scalar field propagating in the higher-dimensional electrically charged regular black hole spacetime.

Following the method proposed by Kanti \textit{et al.} \cite{Kanti:2014dxa}, we divide the spacetime into three regions to facilitate the matching process:
\begin{itemize}
\item \textbf{Region I}: Near the event horizon ($r \approx r_h$), where the effective potential $V(r) \ll \omega^2$.
\item \textbf{Region~II}: The intermediate region between the event horizon and the cosmological horizon, where $V(r) \gg \omega^2$.
\item \textbf{Region~III}: Near the cosmological horizon ($r \approx r_c$), where the effective potential again satisfies $V(r) \ll \omega^2$.
\end{itemize}
We will obtain approximate solutions in Regions I and III and then match them in Region~II to compute the greybody factor.

\subsection{Near the Event Horizon}

To solve the radial equation near the black hole event horizon, we make the following coordinate transformation and redefine the cosmological constant:
\begin{equation}
r \rightarrow f(r) = \frac{h}{1 - \tilde{\Lambda} r^{2}}, \quad \tilde{\Lambda} = \frac{2\Lambda}{(d - 1)(d - 2)},
\label{eq:EventTrans}
\end{equation}
where $h = h(r)$ is the metric function defined earlier. The transformed variable $f$ varies from 0 to 1 as $r$ goes from $r_h$ to $r \gg r_h$. The derivative satisfies the relation
\begin{equation}
\frac{df}{dr} = \frac{1 - f}{r} \frac{A(r)}{1 - \tilde{\Lambda} r^{2}},
\end{equation}
where
\begin{equation}
A(r) = d - 3 + (d - 1) \tilde{\Lambda} r^{2} - \frac{(d - 1)(1 - \tilde{\Lambda} r^{2}) q^{d - 2}}{r^{d - 2} + q^{d - 2}},
\end{equation}
which reduces to $A_{\text{SdS}} = d - 3 + (d - 1) \tilde{\Lambda} r^{2}$ for the Schwarzschild--Tangherlini--de Sitter (SdS) black hole when $q = 0$.

In terms of the transformed variables, the radial equation near the event horizon takes the form
\begin{equation}
\begin{aligned}
&f(1 - f) \frac{d^{2}\phi}{df^{2}} + \left(1 - B_{h} f \right) \frac{d\phi}{df} \\ & + \left[ -\frac{(\omega r_{h})^{2}}{A_{h}^{2}} + \frac{(\omega r_{h})^{2}}{A_{h}^{2} f} - \frac{\lambda_{h} (1 - \tilde{\Lambda} r_{h}^{2})}{A_{h}^{2} (1 - f)} \right] \phi = 0, \label{eq:RadialEqEvent}
\end{aligned}
\end{equation}
where we have used the abbreviations
\begin{align}
B_{h} &= 2 - \frac{1 - \tilde{\Lambda} r_{h}^{2}}{A_{h}^{2}} \left[ (d - 3) A_{h} + r A'(r_{h}) \right], \\
\lambda_{h} & = l (l + d - 3) + \xi R^{(h)} r_{h}^{2},
\end{align}
with $A_{h} = A(r_{h})$, and $R^{(h)}$ is the Ricci scalar evaluated at the event horizon $r = r_{h}$, given by
\begin{equation}
R^{(h)} = -h'' + (d - 2) \frac{ -2 r h' + (d - 3) (1 - h) }{ r^{2} } \Big|_{r = r_{h}}.
\end{equation}
We have also made use of the approximation near the event horizon $f \approx 0$:
\begin{equation}
\frac{ (\omega r_{h})^{2} }{ A_{h}^{2} f (1 - f) } \approx \frac{ (\omega r_{h})^{2} (1 - f) }{ A_{h}^{2} f } = -\frac{ (\omega r_{h})^{2} }{ A_{h}^{2} } + \frac{ (\omega r_{h})^{2} }{ A_{h}^{2} f }.
\end{equation}
This manipulation avoids unphysical behaviour that may arise due to the poles of the Gamma function, as noted in Ref.~\cite{Zhang:2017yfu}.

By redefining the field as $\phi(f) = f^{\alpha_{1}} (1 - f)^{\beta_{1}} W(f)$, Eq.~\eqref{eq:RadialEqEvent} becomes a hypergeometric equation:
\begin{equation}
\begin{aligned}
&f(1 - f) \frac{d^{2} W}{df^{2}} + \left[ 1 + 2 \alpha_{1} - \left( 2 \alpha_{1} + 2 \beta_{1} + B_{h} \right) f \right] \frac{dW}{df}\\
&- \frac{ \omega^{2} r_{h}^{2} + A_{h}^{2} (\alpha_{1} + \beta_{1})(B_{h} + \alpha_{1} + \beta_{1} - 1) }{ A_{h}^{2} } W = 0.
\end{aligned}
\end{equation}
The coefficients are given by
\begin{equation}
\begin{aligned}
\alpha_{1} = &\pm i \frac{ \omega r_{h} }{ A_{h} }, \\
\beta_{1} = & \frac{1}{2} \left( 2 - B_{h} \pm \sqrt{ (2 - B_{h})^{2} + \frac{4 \lambda_{h} (1 - \tilde{\Lambda} r_{h}^{2}) }{ A_{h}^{2} } } \right).
\end{aligned}
\end{equation}

The solution to the above equation is the standard hypergeometric function $F(a_{1}, b_{1}, c_{1}; f)$, with parameters
\begin{align}
a_{1} &= \alpha_{1} + \beta_{1} + \frac{1}{2} \left( B_{h} - 1 + \sqrt{ (1 - B_{h})^{2} - \frac{4 \omega^{2} r_{h}^{2} }{ A_{h}^{2} } } \right), \nonumber \\
b_{1} &= \alpha_{1} + \beta_{1} + \frac{1}{2} \left( B_{h} - 1 - \sqrt{ (1 - B_{h})^{2} - \frac{4 \omega^{2} r_{h}^{2} }{ A_{h}^{2} } } \right), \\
c_{1} &= 1 + 2 \alpha_{1}. \nonumber
\end{align}
Considering the original and redefined fields $\phi(f)$ and $W(f)$ near the event horizon, the radial function $\phi(f)$ takes the form
\begin{equation}
\begin{aligned}
\phi_{H} =& A_{1} f^{\alpha_{1}} (1 - f)^{\beta_{1}} F(a_{1}, b_{1}, c_{1}; f) + A_{2} f^{-\alpha_{1}} (1 - f)^{\beta_{1}}\\ & \times F(1 + a_{1} - c_{1}, 1 + b_{1} - c_{1}, 2 - c_{1}; f),
\end{aligned}
\end{equation}
where $A_{1}$ and $A_{2}$ are constants. In the vicinity of the event horizon, $f \rightarrow 0$, and the solution simplifies to
\begin{equation}\label{eqn42}
\phi_{H} \simeq A_{1} f^{\alpha_{1}} + A_{2} f^{ -\alpha_{1} }.
\end{equation}

At the black-hole horizon, the effective potential vanishes. Consequently, the general solution in this region can be expected to take the form of free plane waves,
\begin{equation}\label{eqn43}
    \phi_H \sim \tilde{A}_1 e^{-i \omega r_\star} + \tilde{A}_2 e^{i \omega r_\star}.
\end{equation}
It is clear that for \( f \propto e^{A_h r_\star / r_h} \), the two solutions \eqref{eqn42} and \eqref{eqn43} are consistent under suitable redefinitions of the integration constants. Since we require only ingoing waves at the event horizon, we choose $\alpha_{1} = -i \omega r_{h} / A_{h}$, corresponding to the ingoing mode, and set $A_{2} = 0$. Additionally, to satisfy the convergence condition (which imples $Re(c_{1}-a_{1}-b_{1})>0$) of the hypergeometric function, we take the negative branch of $\beta_{1}$. Therefore, near the event horizon, the solution becomes
\begin{equation}
\phi_{H} = A_{1} f^{\alpha_{1}} (1 - f)^{\beta_{1}} F(a_{1}, b_{1}, c_{1}; f).
\label{eq:EventSol}
\end{equation}

\subsection{Near the Cosmological Horizon}

We now focus on the radial region near the cosmological horizon $r_c$, employing a similar approach. Here, the metric function $h$ is approximated as \cite{Harmark:2007jy, Crispino:2013pya, Kanti:2014dxa, Zhang:2017yfu}
\begin{equation}
h(r) = 1 - \tilde{\Lambda} r^{2} - \left( \frac{ r_{h} }{ r } \right)^{d - 3} (1 - \tilde{\Lambda} r_{h}^{2} ) \approx \tilde{h} = 1 - \tilde{\Lambda} r^{2},
\end{equation}
valid in the limit where $r \gg r_{h}$. The function $\tilde{h}$ varies from 0 at $r = r_{c}$ to 1 as $r \ll r_{c}$. This approximation is more accurate for small $\tilde{\Lambda}$ (large $r_c$) and higher spacetime dimensions $d$.

We make the change of variable $r \rightarrow \tilde{h}(r)$, so that the radial equation near the cosmological horizon becomes
\begin{equation} \label{Xh}
\begin{aligned}
& \tilde{h} (1 - \tilde{h} ) \frac{ d^{2} \phi }{ d \tilde{h}^{2} } + \left( 1 - \frac{ d + 1 }{ 2 } \tilde{h} \right) \frac{ d\phi }{ d \tilde{h} } \\ &+ \left[ \frac{ (\omega r_{c})^{2} }{ 4 \tilde{h} } - \frac{ l ( l + d - 3 ) }{ 4 (1 - \tilde{h} ) } - \frac{ \xi R^{(c)} r_{c}^{2} }{ 4 } \right] \phi = 0,
\end{aligned}
\end{equation}
where $R^{(c)}$ is the Ricci scalar at $r = r_{c}$, given by
\begin{equation}
R^{(c)} = -\tilde{h}'' + (d - 2) \frac{ -2 r \tilde{h}' + (d - 3)(1 - \tilde{h}) }{ r^{2} } \Big|_{ r = r_{c} }.
\end{equation}
By redefining the field as $\phi( \tilde{h} ) = \tilde{h}^{ \alpha_{2} } (1 - \tilde{h} )^{ \beta_{2} } X( \tilde{h} )$, the above equation becomes a hypergeometric equation
\begin{equation}
\begin{aligned}
&\tilde{h} (1 - \tilde{h} ) \frac{ d^{2} X }{ d \tilde{h}^{2} } + \left[ 1 + 2 \alpha_{2} - \left( 2 \alpha_{2} + 2 \beta_{2} + \frac{ d + 1 }{ 2 } \right) \tilde{h} \right] \frac{ dX }{ d \tilde{h} } \\ & - \frac{ 2 ( \alpha_{2} + \beta_{2} )( \alpha_{2} + \beta_{2} + d - 1 ) + \xi R^{(c)} r_{c}^{2} }{ 4 } X = 0,
\end{aligned}
\end{equation}
with
\begin{equation}
\alpha_{2} = \pm i \frac{ \omega r_{c} }{ 2 }, \quad
\beta_{2} = - \frac{ d + l - 3 }{ 2 } \quad \text{or} \quad \frac{ l }{ 2 }.
\end{equation}
The solution to this hypergeometric equation is
\begin{equation}
\begin{aligned}
 \phi_{C} = &B_{1} \tilde{h}^{ \alpha_{2} } (1 - \tilde{h} )^{ \beta_{2} } F( a_{2}, b_{2}, c_{2}; \tilde{h} ) + B_{2} \tilde{h}^{ -\alpha_{2} } (1 - \tilde{h} )^{ \beta_{2} } \\ & \times F( 1 + a_{2} - c_{2}, 1 + b_{2} - c_{2}, 2 - c_{2}; \tilde{h} ),
\end{aligned}
\end{equation}
with parameters
\begin{align}
a_{2} &= \alpha_{2} + \beta_{2} + \frac{ d - 1 + \sqrt{ ( d - 1 )^{2} - 4 \xi R^{(c)} r_{c}^{2} } }{ 4 }, \\
b_{2} &= \alpha_{2} + \beta_{2} + \frac{ d - 1 - \sqrt{ ( d - 1 )^{2} - 4 \xi R^{(c)} r_{c}^{2} } }{ 4 }, \nonumber \\
c_{2} &= 1 + 2 \alpha_{2}. \nonumber
\end{align}
Here, $B_{1}$ and $B_{2}$ are constants. To satisfy the convergence condition of the hypergeometric function (which implies $Re(c_{2}-a_{2}-b_{2})>0$), we take $\beta_{2} = - ( d + l - 3 ) / 2$.

Near the cosmological horizon $\tilde{h} \rightarrow 0$, the metric function and the effective potential vanish, so the solution represents a superposition of ingoing and outgoing plane waves:
\begin{equation}
\phi_{C} = B_{1} e^{ -i \omega r_{*} } + B_{2} e^{ i \omega r_{*} },
\end{equation}
where $r_{*}$ is the tortoise coordinate near $r = r_{c}$, given by $r_{*} = \frac{1}{2} r_{c} \ln \left( \frac{ r / r_{c} + 1 }{ r / r_{c} - 1 } \right )$. The negative exponent corresponds to an ingoing wave, while the positive exponent corresponds to an outgoing wave. We chose $\alpha_{2} = i \omega r_{c} / 2$ to match the behaviour of the solution. Unlike near the event horizon, both ingoing and outgoing waves exist here. The amplitudes of these waves determine the greybody factor for scalar fields emitted by the black hole:
\begin{equation}
| \gamma_{ \omega l } |^{2} = 1 - \left| \frac{ B_{2} }{ B_{1} } \right|^{2}.
\end{equation}

\subsection{Matching the Solutions in the Intermediate Region}
Having derived asymptotic solutions near both the black hole and cosmological horizons, the solution is only complete if these solutions can be smoothly matched for some radial coordinate value in the intermediate region. This ensures that the full radial solution is well-defined across the entire spacetime.

\subsubsection{Black Hole Horizon}
We begin by considering the solution near the event horizon. In the intermediate region, where $r \gg r_{h}$, the function $f \rightarrow 1$. This limit allows us to rewrite the hypergeometric function from the variable $f$ to $(1 - f)$,
\begin{widetext}
\begin{align}
F(a,b,c;f) &= \frac{\Gamma(c)\Gamma(c - a - b)}{\Gamma(c - a)\Gamma(c - b)} F(a,b,a+b-c+1;1-f) \\
& \quad + (1 - f)^{c - a - b} \frac{\Gamma(c)\Gamma(a+b-c)}{\Gamma(a)\Gamma(b)} F(c-a,c-b,c-a-b+1;1-f). \nonumber
\end{align}
\end{widetext}
Adopting the small-value approximation $\Lambda r_{h}^{2} \ll 1$ and recalling that $A_{h} \simeq d - 3$ in the region where $r \gg r_{h}$, we ensure this approximation remains valid as long as $\Lambda r^{2} \simeq r^{2}/r_{c}^{2} \ll 1$. From Eq.~\eqref{eq:EventTrans}, in the limit $r \gg r_{h}$, we have
\begin{equation}
h \rightarrow 1 - \tilde{\Lambda} r^{2} + \mathcal{O}\left(\frac{r_{h}^{d-3}}{r^{d-3}}\right).
\end{equation}
Under these conditions, the Ricci scalar $R^{(h)} \rightarrow \frac{2d\Lambda}{d-2}$. For small $\xi$, the term $\xi R^{(h)} r_{h}^{2} \rightarrow \xi \frac{2d\Lambda r_{h}^{2}}{d-2} \ll 1$, allowing us to neglect it. Hence, we set $B_{h} \simeq 1$ and $\beta_{1} \simeq -\frac{l}{d-3}$.

From these approximations, we find
\begin{eqnarray}
1 - f & \simeq & \left(1 + \frac{q^{d-2}}{r_{h}^{d-2}}\right)^{\frac{d-1}{d-2}}\left(\frac{r_{h}}{r}\right)^{d-3}, \nonumber \\
\beta_{1} + c_{1} - a_{1} - b_{1} & \simeq & \frac{l + d - 3}{d - 3}.
\end{eqnarray}

In this limit, the solution \eqref{eq:EventSol} becomes
\begin{equation}
\phi_{H} \simeq \Sigma_{2} r^{l} + \Sigma_{1} r^{-l-d+3}, \label{eq:EventMatch}
\end{equation}
where
\begin{equation}
\begin{aligned}
\Sigma_{1} = & A_{1}\frac{\Gamma(c_{1})\Gamma(a_{1}+b_{1}-c_{1})}{\Gamma(a_{1})\Gamma(b_{1})} \\ & \times \left(1+\frac{q^{d-2}}{r_{h}^{d-2}}\right)^{\frac{(d-1)(l+d-3)}{(d-2)(d-3)}}r_{h}^{l+d-3},\\
\Sigma_{2} = & A_{1}\frac{\Gamma(c_{1})\Gamma(c_{1}-a_{1}-b_{1})}{\Gamma(c_{1}-a_{1})\Gamma(c_{1}-b_{1})} \\ & \times\left(1+\frac{q^{d-2}}{r_{h}^{d-2}}\right)^{\frac{-(d-1)l}{(d-2)(d-3)}}r_{h}^{-l}.
\end{aligned}
\end{equation}

As emphasized in Ref.~\cite{Kanti:2014dxa}, the above approximations are strictly valid only for expressions involving the factor $(1-f)$, not for the arguments of the Gamma functions themselves, to maintain the accuracy of the analytical results.

\subsubsection{Cosmological Horizon}
We now consider the asymptotic solution near the cosmological horizon. Similar to the previous case, we switch the argument of the hypergeometric function from $\tilde{h}$ to $(1 - \tilde{h})$ as $\tilde{h} \to 1$. We again use the small-value approximation for the cosmological constant. In the region where $r \ll r_{c}$,
\begin{equation}
1 - \tilde{h} \simeq \left(\frac{r}{r_{c}}\right)^{2}.
\end{equation}
We also have $\beta_{2} \simeq -\frac{l + d - 3}{2}$ and $(\beta_{2} + c_{2} - a_{2} - b_{2}) \simeq \frac{l}{2}$. Following a procedure similar to the event horizon case, we find
\begin{equation}
\phi_{C} \simeq (\Sigma_{3} B_{1} + \Sigma_{4} B_{2}) r^{-(l + d - 3)} + (\Sigma_{5} B_{1} + \Sigma_{6} B_{2}) r^{l}, \label{eq:CosmMatch}
\end{equation}
where
\begin{equation}
\begin{aligned}
\Sigma_{3} &= \frac{\Gamma(c_{2})\Gamma(c_{2}-a_{2}-b_{2})}{\Gamma(c_{2}-a_{2})\Gamma(c_{2}-b_{2})}r_{c}^{l+d-3}, \\  \Sigma_{4} &=\frac{\Gamma(2-c_{2})\Gamma(c_{2}-a_{2}-b_{2})}{\Gamma(1-a_{2})\Gamma(1-b_{2})}r_{c}^{l+d-3},\\[6pt]
\Sigma_{5} &= \frac{\Gamma(c_{2})\Gamma(a_{2}+b_{2}-c_{2})}{\Gamma(a_{2})\Gamma(b_{2})}r_{c}^{-l}, \\ \Sigma_{6} &=\frac{\Gamma(2-c_{2})\Gamma(a_{2}+b_{2}-c_{2})}{\Gamma(a_{2}-c_{2}+1)\Gamma(b_{2}-c_{2}+1)}r_{c}^{-l}.
\end{aligned}    
\end{equation}

Since the two solutions \eqref{eq:EventMatch} and \eqref{eq:CosmMatch} share the same power-law form, matching them is straightforward. Equating coefficients of identical powers of $r$ in Eqs.~\eqref{eq:EventMatch} and \eqref{eq:CosmMatch}, we obtain
\begin{equation}
\Sigma_{3} B_{1} + \Sigma_{4} B_{2} = \Sigma_{1}, \quad \Sigma_{5} B_{1} + \Sigma_{6} B_{2} = \Sigma_{2}.
\end{equation}
Solving for $B_{1}$ and $B_{2}$ and substituting back into the greybody factor expression, we find
\begin{equation}
|\gamma_{\omega l}|^{2} = 1 - \left|\frac{\Sigma_{2}\Sigma_{3} - \Sigma_{1}\Sigma_{5}}{\Sigma_{1}\Sigma_{6} - \Sigma_{2}\Sigma_{4}}\right|^{2}. \label{eq:GammaGrey}
\end{equation}

This result shares the same general form as those found in Einstein gravity \cite{Kanti:2014dxa, Ahmed:2016lou} and Einstein--Gauss--Bonnet gravity \cite{Zhang:2017yfu}. However, in this case, the presence of the non-linear charge $q$, along with the cosmological constant $\Lambda$ and the non-minimal coupling $\xi$, influences the scalar field propagation and, consequently, the greybody factor.

The obtained expression \eqref{eq:GammaGrey} is suitable for small values of $\Lambda$ and when the distance between the two horizons is large. Moreover, since the energy of the emitted particle does not appear explicitly in our approximation, the result suggests validity across all energy ranges. Nevertheless, our findings are most accurate in the low-energy regime. Deviations at higher energies call for further investigation, potentially requiring numerical methods to solve Eq.~\eqref{eq:RadialEq}.

\begin{figure*}[t]
\centering
\includegraphics[width=\textwidth]{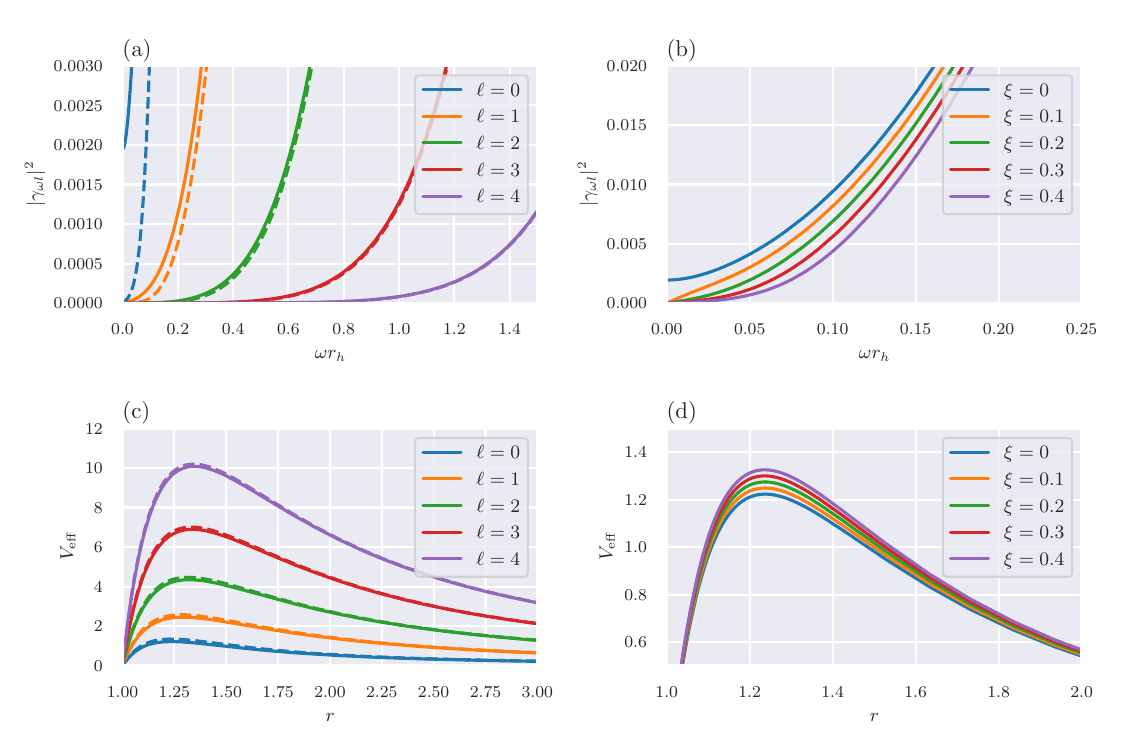}
\caption{Effect of particle properties on the greybody factor (upper panels) and the enhancement/suppression of the corresponding effective potential (lower panels) for $d=6$, $\Lambda =0.01$, and $q=0.5$. The left panels show $l=0,1,2,3,4$ for $\xi=0$ (solid lines) and $\xi=0.5$ (dashed lines), while the right panels show $l=0$ with $\xi =0,0.1,0.2,0.3,0.4$.}
\label{fig1}
\end{figure*}

\subsection{Low-Energy Limit}
Having established that our analytic expression provides a correct profile in the low-energy regime, we now derive the low-energy limit ($\omega \rightarrow 0$) for both minimally and non-minimally coupled scalar fields.

For the minimal coupling case ($\xi = 0$) and the lowest mode of propagation ($l = 0$), one can show that
\begin{equation}
\begin{aligned}
&\Sigma _1 \sim A_1 \frac{i\omega}{(2 - B_{h0}) A_{h0}} r_h^{d - 3} \left[\left(\frac{q}{r_h}\right)^{d - 2} + 1\right]^{\frac{d - 1}{d - 2}} + \mathcal{O}(\omega^{2}), \\
&\Sigma_{2} \sim A_1 + \mathcal{O}(\omega),\\
&\Sigma _3 \sim \frac{i\omega}{d - 3}r_c^{d - 2} + \mathcal{O}(\omega^{2}), \ \Sigma _4 \sim \frac{-i\omega}{d - 3} r_c^{d - 2} + \mathcal{O}(\omega^{2}), \\ &\Sigma _{5,6} \sim 1 + \mathcal{O}(\omega).
\end{aligned}
\end{equation}
Here,
\begin{equation*}
A_{h0} = (d - 3) - \frac{(d - 1) q^{d - 2}}{q^{d - 2} + r_h^{d - 2}},
\end{equation*}
and
\begin{equation*}
\begin{aligned}
B_{h0} = & \frac{1}{\left((d - 3) q^{2} r_h^{d} - 2 r_h^{2} q^{d}\right)^{2}} \\ & \times \left[ \right. \left(-2 d^{2} + 3 d + 7\right) q^{d + 2} r_h^{d + 2}  \left. + 2(d + 1) r_h^{4} q^{2d} \right. \\ & \left. + (d - 3)^{2} q^{4} r_h^{2d} \right].
\end{aligned}
\end{equation*}
Substituting these into the greybody factor expression yields
\begin{widetext}
\begin{equation}
|\gamma_{\omega l}|^{2} = \frac{4 A_{h0} (2 - B_{h0})(d - 3) r_c^{d + 2} r_{h}^{d + 3} \left(\frac{r_{h}^{2} \left(\frac{q}{r_{h}}\right)^{d}}{q^{2}} + 1\right)^{\frac{d + 1}{d - 2}}}{\left[A_{h0}(B_{h0}-2)r_{h}^{3}r_{c}^{d}\left(\frac{r_{h}^{2} \left(\frac{q}{r_{h}}\right)^{d}}{q^{2}}+1\right)^{\frac{1}{d-2}} - (d - 3) r_{c}^{2} r_{h}^{d}\left(\frac{r_{h}^{2} \left(\frac{q}{r_{h}}\right)^{d}}{q^{2}}+1\right)^{\frac{d}{d - 2}}\right]^{2}} + \mathcal{O}(\omega).
\end{equation}
\end{widetext}

This shows that the probability of emitting low-energy scalar particles remains non-zero, even for a higher-dimensional electrically charged regular black hole. Thus, the fundamental feature of scalar propagation in a de Sitter background is preserved. At the same time, the presence of the non-linear charge $q$ significantly modifies the greybody factor.

In the non-minimal coupling case ($\xi \neq 0$), evaluating the limiting forms becomes more challenging. However, in the low-energy regime, one can still demonstrate that
\begin{equation}
\begin{aligned}
& \Sigma_{2}\Sigma_{3} = E + i \Sigma_{231}\omega + \Sigma_{232}\omega^{2}, \\
& \Sigma_{1}\Sigma_{5} = K + i\Sigma_{151}\omega + \Sigma_{152}\omega^{2}, \\
& \Sigma_{2}\Sigma_{4} = E + i\Sigma_{241}\omega + \Sigma_{242}\omega^{2}, \\
& \Sigma_{1}\Sigma_{6} = K + i\Sigma_{161}\omega + \Sigma_{162}\omega^{2}.
\end{aligned}
\end{equation}

From these expansions, it follows that
\begin{equation}
|\Sigma_{2}\Sigma_{3} - \Sigma_{1}\Sigma_{5}|^{2} \sim |\Sigma_{1}\Sigma_{6} - \Sigma_{2}\Sigma_{4}|^{2} = (K - E)^{2} + \mathcal{O}(\omega^{2}).
\end{equation}
Substituting this back into the greybody factor expression, we obtain a non-zero term for low-energy scalar emission in the non-minimal coupling case. This result holds for arbitrary modes, indicating that non-minimal coupling effectively removes scalar modes exhibiting non-zero low-energy asymptotic greybody factors.

\begin{figure*}[t]
\centering
\includegraphics[width=\textwidth]{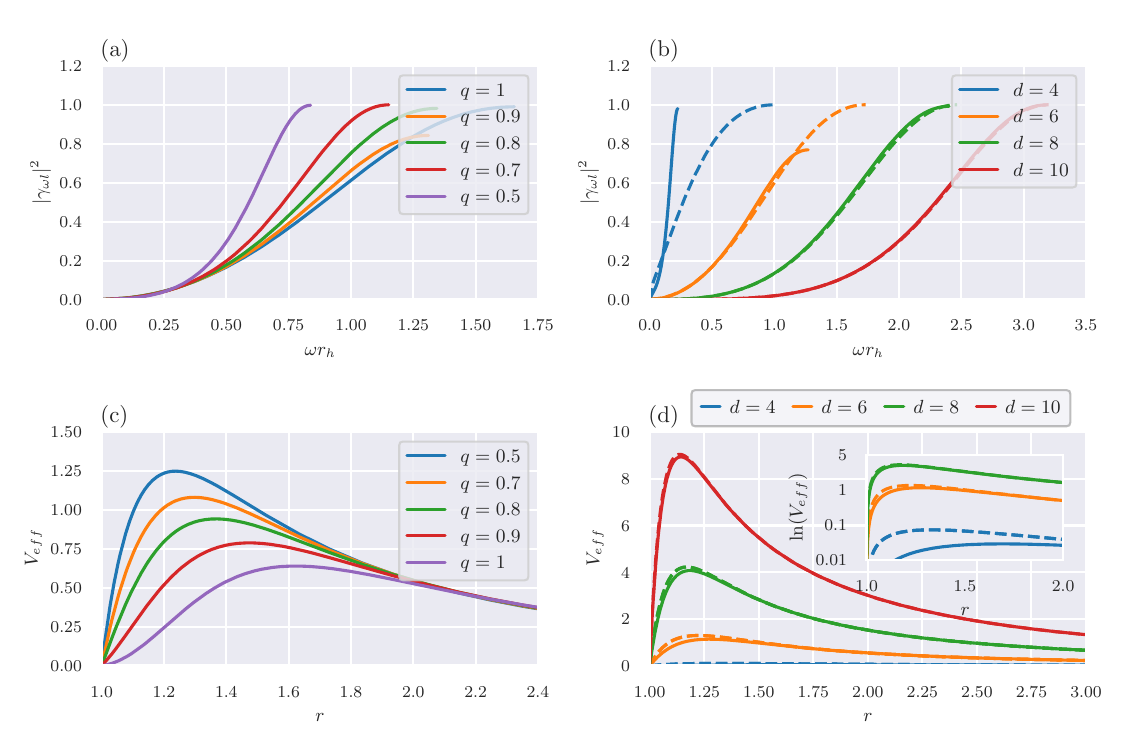}
\caption{Effect of spacetime properties $(q, d)$ on the greybody factor (upper panels) and the enhancement/suppression of the corresponding effective potential (lower panels) for $\Lambda=0.01$, $l=0$, and $\xi=0.1$. Left panels: $q=0.5,0.7,0.8,0.9,1$ for $d=6$. Right panels: $d=4,6,8,10$ with $q \to 0$ (dashed lines) and $q=0.65$ (solid lines). The inset of Fig. 2(d) shows a magnified view of the effective potential for $d=4$.}
\label{fig2}
\end{figure*}
\section{The Effect of Spacetime and Particle Properties on the Greybody Factor} \label{section5}

The propagation of scalar fields in the gravitational background of a charged, regular de Sitter black hole is influenced by both spacetime and particle parameters. The particle properties, specifically the angular momentum number $l$ and the non-minimal coupling constant $\xi$, play significant roles. Similarly, the cosmological constant $\Lambda$ and the spacetime dimension $d$ affect propagation in a manner analogous to the SdS case. Here, we have an additional spacetime parameter: the non-linear coupling $q$, on which we focus more closely. We investigate the dependence of the greybody factor on these parameters, alongside the corresponding changes in the effective potential.

We consider the greybody factor $|\gamma_{\omega l}|^2$ as a function of the dimensionless energy parameter $\omega r_h$. To gain an intuitive understanding of particle propagation, we present the effective potential below the greybody factor plots.

Figure~\ref{fig1} illustrates the effect of particle parameters $(l, \xi)$. In the left panel, we show the first five partial waves for $l=0,1,2,3,4$ with $\Lambda =0.01$ (in units of $r_h^{-2}$). The solid lines correspond to the minimal coupling case $\xi=0$, and the dashed lines correspond to the non-minimal coupling case $\xi=0.5$. For both cases, increasing $l$ suppresses the greybody factor, as higher modes face higher potential barriers. Regardless of whether the scalar field is minimally or non-minimally coupled, the dominant mode remains $l=0$. Nevertheless, the coupling parameter $\xi$ decreases the greybody factor for all modes, with the effect being most pronounced for the dominant mode. This influence of $\xi$ is also evident in the effective potential. For $\xi=0$, the low-energy greybody factor is non-zero, as shown previously; however, non-minimal coupling alters this feature. In fact, in the right-hand panel of Fig.~\ref{fig1}, we study the dependence of particle propagation on $\xi$ for the lowest mode $l=0$. Since non-minimal coupling reduces the effective potential, the corresponding greybody factor is enhanced.

\begin{figure*}[t]
\centering
\includegraphics[width=\textwidth]{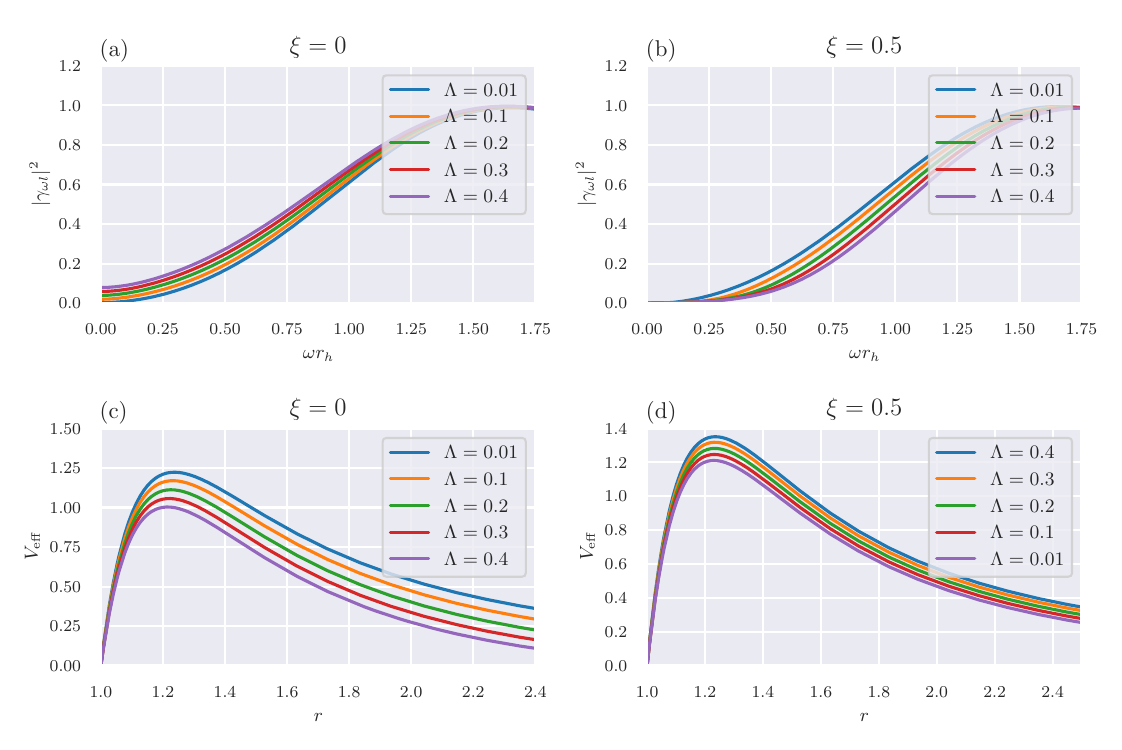}
\caption{Effect of cosmological constant $\Lambda$ on the greybody factor (upper panels) and the corresponding effective potential (lower panels) for $d=6$, $l=0$, $q=0.5$, and $\Lambda=0.01,0.1,0.2,0.3,0.4$. The left panels correspond to $\xi=0$, and the right panels to $\xi=0.5$.}
\label{fig3}
\end{figure*}

Next, we focus on the effect of spacetime properties on the greybody factor, although we discuss the role of the cosmological constant separately due to its peculiar behaviour. In Fig.~\ref{fig2}, we present the roles of $(q,d)$ for the dominant mode. The non-linear charge $q$ enhances the greybody factor by significantly lowering the effective potential. The greybody factors for various spacetime dimensions are shown both with and without a non-linear charge ($q \to 0$). For higher dimensions, the greybody factor is suppressed, as the potential barrier increases considerably. While the non-linear charge modifies the saturation point of the greybody factor, it must still approach 1 in all cases at high energies, as high-energy modes easily overcome the potential barrier. We plot the greybody factors in the low-energy region, since the matching approach employed in deriving $\gamma_{\omega l}$ is valid in this regime. Moreover, our analytical approach is limited to even-dimensional spacetimes due to the appearance of Poles in gamma functions, which constrains our ability to evaluate the greybody factor. This issue has been observed in previous works \cite{Kanti:2014dxa}.

Finally, we examine the influence of the cosmological constant in Fig.~\ref{fig3}. Notably, its effect depends on the non-minimal coupling parameter $\xi$. For small $\xi$, increasing $\Lambda$ enhances the greybody factor, whereas beyond a certain threshold in $\xi$, it suppresses scalar emission. This competition between $\xi$ and $\Lambda$ is also reflected in the effective potential. The dual role played by $\Lambda$, as a source of homogeneously distributed energy (enhancing effect) and as an effective mass term via non-minimal coupling (suppressing effect), leads to this intricate interplay. Similar phenomena have been observed in SdS and Gauss-Bonnet dS spacetimes.

\section{Power Spectra of Hawking Radiation} \label{secgion6}
Hawking radiation can be intuitively understood as a virtual particle-antiparticle pair spontaneously forming just outside the black hole horizon. In this picture, the particle with positive energy escapes to infinity, while its negative-energy counterpart is drawn into the black hole. Once inside, the negative-energy particle can be interpreted as an ordinary particle, and this process effectively reduces the black hole mass. The resulting emission observed at infinity is thermal, characterized by the Hawking temperature $T_{H}$. However, the actual spectrum is not a perfect blackbody due to frequency-dependent transmission probabilities, known as greybody factors, which modify the pure thermal emission.

\begin{figure*}[t]
\centering
\includegraphics[width=\textwidth]{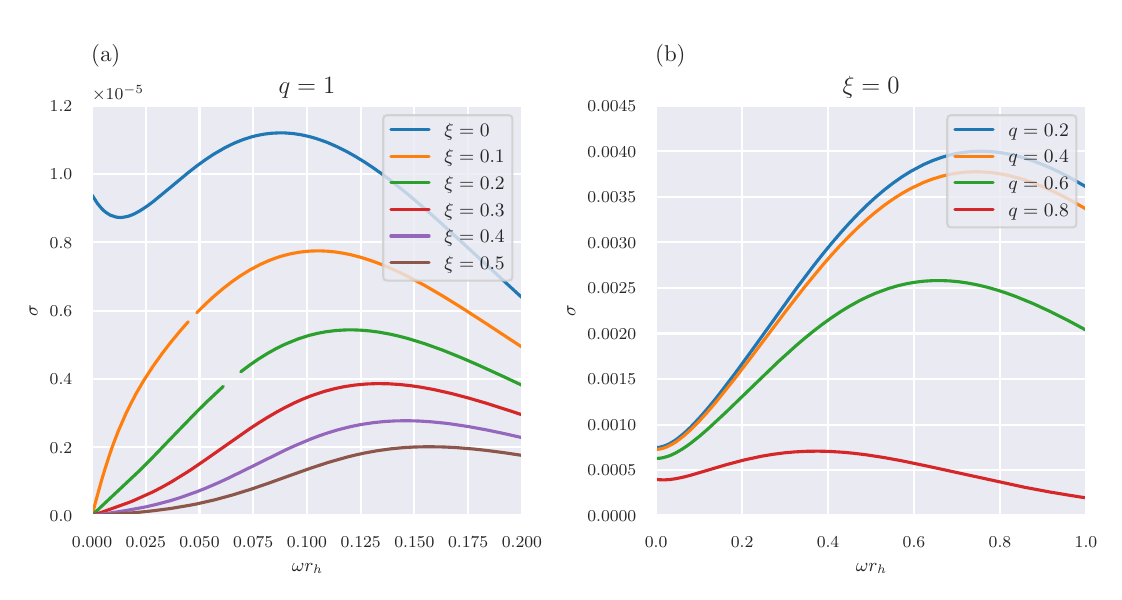}
\caption{Power spectra of Hawking radiation for $d=6$, $\Lambda=0.01$. Left panel: $q=1$ with $\xi=0,0.1,0.2,0.3,0.4,0.5$. Right panel: $\xi=0$ with $q=0.2,0.4,0.6,0.8$.}
\label{fig4}
\end{figure*}

For a higher-dimensional spherically symmetric black hole described by the metric (\ref{eq:4metric}), the flux spectrum (the number of emitted particles per unit time) for massless scalar particles generalizes the four-dimensional result \cite{Hawking:1974sw} to $d$ dimensions,
\begin{equation}
\label{flux}
\frac{dN(\omega)}{dt} = \sum_{\ell} \sigma_{\ell,d}(\omega) \frac{1}{\exp(\omega/T_{H}) - 1} \frac{d^{d-1}\omega}{(2\pi)^{d-1}}.
\end{equation}
The power spectrum (energy emitted per unit time) is obtained by incorporating the energy carried by each particle,
\begin{equation}
\frac{dE(\omega)}{dt} = \sum_{\ell} \sigma_{\ell,d}(\omega) \frac{\omega}{\exp(\omega/T_{H}) - 1} \frac{d^{d-1}\omega}{(2\pi)^{d-1}}.
\label{power}
\end{equation}
The factor $\sigma_{\ell,d}(\omega)$, which differs from unity, introduces corrections to the pure blackbody spectrum. It can be determined by solving the field equations for the particle and computing its absorption probability $\gamma_{\omega \ell}$. In fact, one can write \cite{Gubser:1997yh}
\begin{equation}
\begin{aligned}
\sigma_{\ell,d} (\omega) = & \frac{2^{d-4}\pi^{(d-3)/2}\,\Gamma[(d-3)/2]}{(d-4)!\,\omega^{d-2}} \\ & \frac{(2\ell+d-3)(\ell+d-4)!}{\ell!} |\gamma_{\omega \ell}|^2.
\label{grey-n}
\end{aligned}
\end{equation}
A more convenient form is
\begin{equation}
\sigma_{\ell,d} (\omega) = \frac{2^{d-4}}{\pi} \Gamma\Bigl(\frac{d-1}{2}\Bigr)^2 \frac{A_h}{(\omega r_h)^{d-2}} N_\ell |\gamma_{\omega \ell}|^2,
\label{greyb}
\end{equation}
where $N_\ell$ is the multiplicity of states for a given partial wave $\ell$ in $d$ dimensions \cite{Bander:1965rz},
\begin{equation}
N_\ell= \frac{(2\ell+d-3)(\ell+d-4)!}{\ell!(d-3)!},
\label{bulk-mult}
\end{equation}
and $A_h$ is the horizon area of the $d$-dimensional black hole,
\begin{equation}
A_h = r_h^{d-2}(2\pi)\pi^{(d-3)/2}\Gamma\Bigl(\frac{d-1}{2}\Bigr)^{-1}.
\label{area-n}
\end{equation}
Rewriting Eq. (\ref{power}) in terms of the greybody factor yields \cite{Kanti:2004nr, Kanti:2005ja, Bousso:1996au}
\begin{equation}
\frac{d E(\omega)}{dt} = \sum_{\ell} N_\ell |\gamma_{\omega \ell}|^2 \frac{\omega}{\exp(\omega/T_{BH}) - 1}\frac{d\omega}{2\pi}.
\label{alter-bulk}
\end{equation}
We will use this last expression to analyse the energy emission rate from the black hole. 

The Hawking temperature $T_{H}$ is the normalized black hole temperature related to the surface gravity \cite{Kanti:2017ubd,Bousso:1996au}, given by
\begin{equation}
    T_{H} = \frac{1}{\sqrt{h(r_0)}} \frac{q r_h^d (d^2 - 5 d - 2 \Lambda r_h^2 + 6) - 2 (d - 2) r_h^2 q^d}{4 \pi (d - 2) r_h (r_h^2 q^d + q r_h^d)},
\end{equation}
where $\sqrt{h(r_0)}$ is the normalization factor, and $h(r_0)$ is the value of the metric function at its global maximum $r_0$, determined by $h'(r)=0$.

We now study the effects of $\xi$ and $q$ on the power spectra, focusing only on lower partial modes, since higher $l$ values contribute significantly less and are suppressed by orders of magnitude. Figure \ref{fig4} shows the power spectra of Hawking radiation for $d=6$ and $\Lambda=0.01$, varying $q$ and $\xi$. In the left panel, we set $q=1$ and vary $\xi=0,0.1,0.2,0.3,0.4,0.5$. It is clear that $\xi$, with other parameters fixed, reduces the Hawking radiation—consistent with the observed behaviour of the greybody factor in the previous section. In the right panel, we set $\xi=0$ and vary $q=0.2,0.4,0.6,0.8$. Despite earlier results suggesting that $q$ enhances particle propagation, here the effect on the power spectrum is suppressed. This is due to the dependence of the normalized temperature on $q$, which, taken together, leads to diminished energy emissions.

\begin{figure*}[t]
\centering
\includegraphics[width=\textwidth]{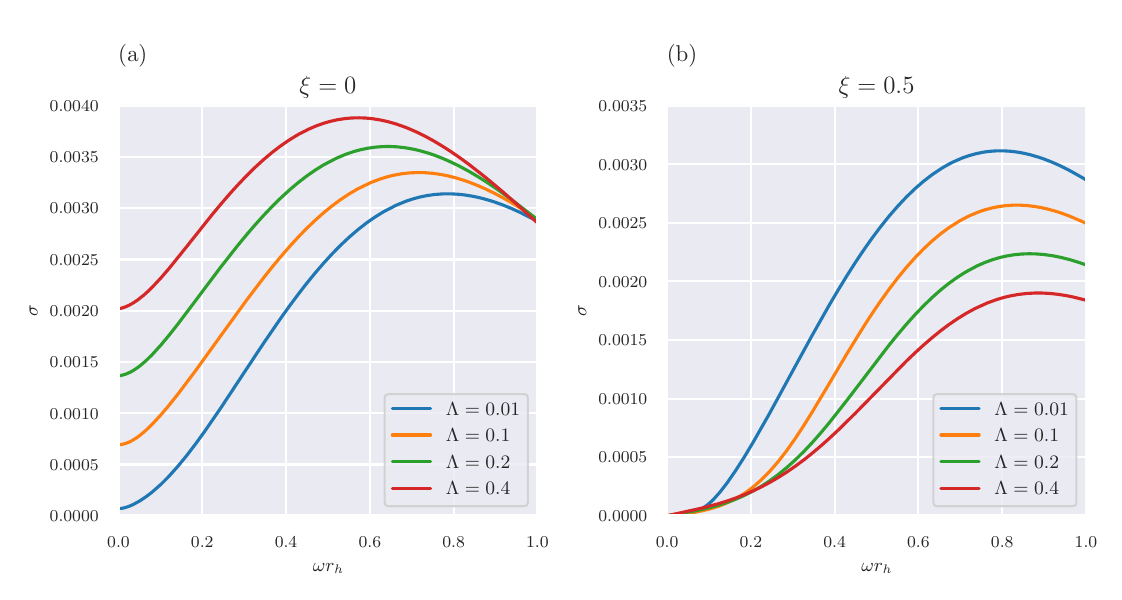}
\caption{Power spectra of Hawking radiation for $d=6$, $q=0.5$ with varying $\Lambda=0.01,0.1,0.2,0.4$. The left panel corresponds to $\xi=0$, while the right panel corresponds to $\xi=0.5$.}
\label{fig5}
\end{figure*}

Finally, we examine the role of the cosmological constant in the energy emission process. Figure \ref{fig5} shows that the interplay between $\xi$ and $\Lambda$, previously observed in the greybody factor, also affects the power spectra. For small $\xi$, increasing $\Lambda$ enhances the Hawking radiation, while for larger $\xi$ values, it suppresses emission. Unlike the greybody factor, the non-linear charge $q$ does not alter this competition between $\xi$ and $\Lambda$.

\section{Conclusion and Discussion}\label{conclusion}

In this work, we analysed the propagation of scalar particles emitted by a class of higher-dimensional, electrically charged, regular black holes in a de Sitter (dS) spacetime. By employing an approximate analytical technique that matches solutions near the event and cosmological horizons, we derived an expression for the greybody factor. We also investigated the low-energy limit of the greybody factor for both minimal and non-minimal couplings of the scalar field. For the minimal coupling scenario, the low-energy asymptotic limit of the greybody factor remains non-zero, consistent with the behaviour of free massless scalar fields in a dS black hole background. In contrast, for the non-minimal coupling case, the same low-energy limit vanishes, indicating that the coupling to curvature alters the characteristic emission profile.

We examined the greybody factor alongside the effective potential to understand how various spacetime and particle parameters influence scalar field propagation. Our findings show that the angular momentum number $l$ and the non-minimal coupling parameter $\xi$ generally suppress the greybody factor. Spacetime properties, however, have more intricate effects: while the non-linear charge parameter $q$ enhances the greybody factor, increasing the spacetime dimension $d$ reduces it. The cosmological constant $\Lambda$ exhibits a dual behaviour-enhancing emission under certain conditions and suppressing it under others, reflecting a non-trivial interplay between $\xi$ and $\Lambda$. Our analysis is primarily valid in the low-energy regime and for arbitrary particle modes. At higher energies, the accuracy of the analytical matching approach diminishes, suggesting that a fully numerical treatment would be beneficial. 

In scenarios involving higher-dimensional black holes, Hawking radiation offers a distinct observational signature: a thermal spectrum that, while resembling a blackbody, is modified by greybody factors. These modifications are crucial, since they alter the emission characteristics—particularly at low and intermediate energies, where particle production is most significant—and thus impact the experimental identification of a black hole event. Higher-dimensional black holes produce radiation both in the bulk and on the brane. While gravitons and certain scalar fields may escape into the bulk, other fields (such as fermions and gauge bosons) remain confined to the brane. Precisely quantifying the energy lost into the bulk is vital for predicting the observed brane emission rates and, ultimately, for designing experiments to detect such phenomena.

Beyond this study, several natural extensions and future directions present themselves. In the asymptotic limit, where the frequency grows large along the imaginary axis, universal results for greybody factors remain elusive. Nonetheless, examining this regime may provide insights into the underlying microscopic structure of black holes, suggesting that their high-frequency behaviour could reflect simpler or more exotic quantum degrees of freedom depending on the spacetime background \cite{Harmark:2007jy}. Investigations in AdS (Anti-de Sitter) backgrounds, as well as other asymptotic geometries, would broaden the applicability of our results. We also note that understanding Hawking radiation from small black holes can serve as a probe of the underlying theory, potentially revealing the dimensionality of spacetime and the presence of large extra dimensions \cite{Kanti:2004nr}.

As a final note, our current focus on scalar fields can be extended to include scalar, spinor, and vector fields, both massless and massive. Examining how different spin fields interact with the gravitational and electromagnetic backgrounds will lead to a more comprehensive understanding of the Hawking radiation spectrum. Such a program would offer a complete picture of emission processes from $d$-dimensional spherically symmetric black holes and yield valuable insights into the fundamental nature of gravity, spacetime dimensions, and field dynamics in curved backgrounds.


\section*{Acknowledgements}
Authors N.K.A., K.H., and A.R.C.L. would like to thank U.G.C. Govt. of India for financial assistance under UGC-NET-SRF scheme. M.S.A.'s research was supported by the ISIRD grant 9-252/2016/IITRPR/708. The research of MSA is supported by the National Postdoctoral Fellowship of the Science and Engineering Research Board (SERB), Department of Science and Technology (DST), Government of India, File No. PDF/2021/00349.

\bibliographystyle{spphys}       
\bibliography{BibTex}
\nocite{}
\end{document}